\documentclass{article}
\usepackage{graphicx}
\usepackage{amsfonts}

\def\ben{\begin{equation}}
\def\een{\end{equation}}

\def\bea{\begin{eqnarray}}
\def\eea{\end{eqnarray}}

\newcommand{\laeq}[1] {\label{eq:#1}}
\newcommand{\reeq}[1] {(\ref{eq:#1})}

\input amssym.def
\input amssym.tex

\begin{document}

\hfuzz=100pt
\title{Symmetric Affine Theories and Nonlinear Einstein-Proca System }
\author{ Miroslav Hejna\footnote{mh403@cam.ac.uk}
\\
DAMTP,
\\ University of Cambridge,
\\Wilberforce Road,
\\Cambridge CB3 0WA, U.K. 
}
\maketitle

%\centerline{\bf {$\frak{D}\frak{R}\frak{A}\frak{F} \frak {T}$}}

\begin{abstract}
We review the correspondence between symmetric affine theories and
the nonlinear Einstein-Proca system that was found by Einstein and
Schr\"odinger. With the use of this correspondence, we investigate
static spherically symmetric solutions in symmetric affine theory
with no matter fields. Use of the correspondence leads to a
significant simplification of the calculation. A special instance
of  ``no-hair'' theorem  for symmetric affine theory is established. 

\end{abstract}

{\small We use reduced Planck units  in which
$c=1$, $\hbar=1$, $8\pi G=1$, $\epsilon_0=1$.
}

\section{Introduction}
The usual Einstein-Hilbert formulation of General Relativity has
several disadvantages. The Hilbert Lagrangian contains second
derivatives of $g_{ab}$, yet it gives second order field equations
(as  opposed to 4-th order that one would expect). This is because
of rather exceptional cancellations of higher order terms. It has
been shown \cite{Kijowski1,Kijowski2} that symmetric affine
theories provide a natural remedy, that considerably clarifies the
canonical structure of General Relativity. Symmetric affine
theories introduce a nonmetric connection, which is desirable as
recent developments suggest that Riemannian description is not
valid on all scales \cite{Fradkin}.

We will review symmetric affine theory and show its equivalence to
the nonlinear Einstein-Proca system. Although the symmetric affine
formalism is presented as more fundamental, certain calculations
may be easier in the corresponding Einstein-Proca system.
Exploiting the equivalence, we show a  version of the ``no-hair''
theorem for symmetric affine theory in spherically symmetric
spacetimes.

\section{Symmetric Affine Theories}

Symmetric affine theories assume that Lagrangian density
$\mathfrak{L}$ is a function of the connection
$\Gamma_{a\phantom{a}c}^{\phantom{a}b} $ and its derivatives. It
is further assumed that the connection is symmetric in its lower
indices. This approach introduces more freedom as there are $40$
independent components of connection, as opposed to $10$
independent components of the metric.

It is reasonable to assume that the Lagrangian  depends on the
connection only through the Riemann tensor $R^a{}_{bcd}$, or
restrict to an even smaller class of Lagrangians that depend only
on the Ricci tensor $R_{ab}=R^e{}_{aeb}$.\footnote{Using
coordinates in which $\Gamma_{a\phantom{a}c}^{\phantom{a}b}=0 $ it
can be easily seen that $R^a{}_{bcd}$ is the only independent
tensor that
 depends on connection and its first derivatives, and is linear in the first derivatives. }
For a general symmetric connection $R_{ab}$ is not necessarily
symmetric, but the first Bianchi identity $R^a{}_{[bcd]}=0$
implies $R^e{}_{eab}=2R_{[ab]}$ and so the only independent
contraction of $R^a{}_{bcd}$ is $R_{ab}=R^e{}_{aeb}.$ (This gives
a partial justification for why we consider only Lagrangians that
depend only on $R_{ab}$). We write $R_{ab}=S_{ab}+F_{ab}$, where
$S_{ab}$ is symmetric and $F_{ab}$ is antisymmetric. The second
Bianchi identity $R^a{}_{b[cd;e]}=0$ then implies $F_{[ab;c]}=0,$
i.e. $F={\rm d}A$.

Symmetric affine theories consider the connection as a basic
concept. The metric density is a derived concept defined as

\ben
\mathfrak{g}^{ab}=2\frac{\partial\mathfrak{L}}{\partial
S_{ab} }= \frac{\partial\mathfrak{L}}{\partial R_{ab}}+
\frac{\partial\mathfrak{L}}{\partial R_{ba}}\,.
\een

We define also
\ben
\mathfrak{G}^{ab}=-2\frac{\partial\mathfrak{L}}{\partial F_{ab} }=
-\frac{\partial\mathfrak{L}}{\partial R_{ab}}+
\frac{\partial\mathfrak{L}}{\partial R_{ba}}\,,
 \een

and current
\ben
\mathfrak{O}^a=-\mathfrak{G}^{ab}{}_{;b}=-\mathfrak{G}^{ab}{}_{,b}\,.
\een

Clearly $\mathfrak{O}^a{}_{,a}=0$, i.e. the current is conserved.

Variation of the action $S=\int \mathfrak{L}(R_{ab})\, d^n x$ and
the use of Palatini lemma leads to the field equations

\ben
2\mathfrak{g}^{ab}{}_{;c}-(\mathfrak{O}^a+\mathfrak{g}^{ad}{}_{;d})\delta^b_c
-(\mathfrak{O}^b+\mathfrak{g}^{bd}{}_{;d})\delta^a_c=0\,.
\een

This is equivalent to

\ben
\mathfrak{g}^{ab}{}_{;c}+\frac{1}{n-1}\mathfrak{O}^a\delta^b_c
+\frac{1}{n-1}\mathfrak{O}^b\delta^a_c=0\,,
\een

where semicolon denotes covariant derivative with connection
$\Gamma_{a\phantom{a}c}^{\phantom{a}b} $.

\pagebreak

Given the metric, there is a unique symmetric metric-compatible
\linebreak connection  $\left\{{}_{a\phantom{a}c}^{\phantom{a}b}
\right\}$ (Levi-Civita connection). It is given by

\ben \left\{{}_{a\phantom{a}c}^{\phantom{a}b}
\right\}=\frac{1}{2}g^{be}(g_{ae,c}+g_{ec,a}-g_{ac,e})\,, \een
 where
$g_{ab}$ is the dedensitized metric and $g^{ab}$ is the inverse
metric.

It follows from the definition of covariant derivative and metric
compatibility of $\left\{{}_{a\phantom{a}c}^{\phantom{a}b}
\right\}$, that the nonmetricity tensor
$Z_{a\phantom{a}c}^{\phantom{a}b}
=\Gamma_{a\phantom{a}c}^{\phantom{a}b}-\left\{{}_{a\phantom{a}c}^{\phantom{a}b}
\right\}$ satisfies

\ben \mathfrak{g}^{ab}{}_{;c}
=Z_{e\phantom{e}c}^{\phantom{e}a}\mathfrak{g}^{eb}+
Z_{e\phantom{e}c}^{\phantom{e}b}\mathfrak{g}^{ea}-
Z_{e\phantom{e}c}^{\phantom{e}e}\mathfrak{g}^{ab}\,.
\een

The equation of motion then gives

\ben
Z_{e\phantom{e}c}^{\phantom{e}a}{g}^{eb}+
Z_{e\phantom{e}c}^{\phantom{e}b}{g}^{ea}-
Z_{e\phantom{e}c}^{\phantom{e}e}{g}^{ab}+
\frac{1}{n-1}{O}^a\delta^b_c
+\frac{1}{n-1}{O}^b\delta^a_c=0
\,,\qquad \laeq{sys}
\een

from which we obtain by contraction
$Z_{e\phantom{a}c}^{\phantom{a}e}=\frac{2}{(n-1)(n-2)}O_a$,

where $g_{ab}$ is used to lower indices.

The system of equation \reeq{sys} can be solved for
$Z_{a\phantom{a}c}^{\phantom{a}b}$. This gives

\ben
\Gamma_{b\phantom{b}c}^{\phantom{b}a}=\left\{{}_{b\phantom{b}c}^{\phantom{b}a}   \right\}-
\frac{1}{n-2}g_{bc}O^a+\frac{1}{(n-1)(n-2)}\left(\delta^a_b O_c
+\delta^a_c O_b\right)\,,\qquad \laeq{connection}
\een

and its contraction

\ben
\Gamma_{a\phantom{a}b}^{\phantom{a}b}=({\ln{\sqrt{g}}})_{,a}+\frac{2}{(n-1)(n-2)}O_a\,.
\een

Working in a frame in which
$\left\{{}_{a\phantom{a}c}^{\phantom{a}b}   \right\}=0 $ and using
$O^a_{\phantom{a};a}=0$, one can easily find an expression for
$R^{\mathrm M}{}_{ab}$, the Ricci tensor arising from metric
$g_{ab}$

$$R_{ab}=R^{\mathrm M}{}_{ab}-\frac{1}{(n-1)(n-2)}O_a O_b+
\frac{1}{(n-1)(n-2)}(O_{b,a}- O_{a,b}).
$$

Comparing to $R_{ab}=S_{ab}+F_{ab}$, we obtain

\ben
S_{ab}=R^{\mathrm M}{}_{ab}-\frac{1}{(n-1)(n-2)}O_a O_b\,,\qquad \laeq{s}
\een

\ben
 F_{ab}=\frac{1}{(n-1)(n-2)}(O_{b,a}- O_{a,b})\,.\qquad
\laeq{f}
\een

The interpretation of these equations is that the presence of
matter fields gives rise to nonmetricity of the connection and
nonmetricity of the Ricci tensor, according to equations
\reeq{connection}, \reeq{s} and \reeq{f}.

\pagebreak

\section{Equivalence of Symmetric Affine Theory and Nonlinear Einstein-Proca Theory}
There is a correspondence between symmetric affine theories and
nonlinear Einstein-Proca theories.

To see this we perform a Legendre transform of the original
Lagrangian

\ben
\bar{\mathfrak{L}}(\mathfrak{g}^{ab},F_{ab})=\mathfrak{L}(S_{ab},F_{ab})
-\frac{1}{2}\mathfrak{g}^{ab}S_{ab}\,.
\een

$S_{ab}$ is eliminated in favour of $\mathfrak{g}^{ab}$ using the
definition
$\mathfrak{g}^{ab}=2\frac{\partial\mathfrak{L}}{\partial S_{ab}
}(S_{ab}, F_{ab}  ) $.

Standard properties of the Legendre transform give

\ben \frac{\partial\bar{\mathfrak{L}}}{\partial F_{ab}
}=\frac{\partial\mathfrak{L}}{\partial F_{ab} }\,,
 \een

\ben \frac{\partial\bar{\mathfrak{L}}}{\partial \mathfrak{g}^{ab}
} =-\frac{1}{2}S_{ab}\,.
\een

We have
$\mathfrak{G}^{ab}=-2\frac{\partial\bar{\mathfrak{L}}}{\partial
F_{ab}}$ and thus we can identify $\mathfrak{G}^{ab}$ with the
Ampere tensor.
We introduce the rescaled current $A_a=\frac{1}{(n-1)(n-2)}O_a$.
Equations \reeq{s}, \reeq{f} can then be rewritten as

\ben
R^{\mathrm M}{}_{ab}=-2\frac{\partial\bar{\mathfrak{L}}}{\partial \mathfrak{g}^{ab} }
+{(n-1)(n-2)}A_a A_b\,,\qquad \laeq{ep1}
\een

\ben
{G}^{ab}{}_{;b}=-(n-1)(n-2)A^a\,,\qquad \laeq{ep2}
\een

\ben
F_{ab}=A_{b,a}- A_{a,b}\,.
\een

Equations \reeq{ep1}, \reeq{ep2},  are precisely the nonlinear
Einstein-Proca equations, which can be derived from the Lagrangian

\ben
{L}_{\mathrm EP}=\frac{1}{2}R^{\mathrm M}{}+\bar{{L}}
-\frac{m_V^2}{2}A^aA_a\,,
\een

where the mass of the vector boson is $m_V^2=(n-1)(n-2).$

Note that if $n>2$, the mass term always gives a negative
contribution to the Raychaudhuri equation. Thus the mass term
always has an attractive effect.
From the Einstein-Proca Lagrangian we see that
$-\bar{\mathfrak{L}}(\mathfrak{g}^{ab},0)$ plays the role of the
cosmological constant density.

Conversely, given a Lagrangian $\bar{\mathfrak{L}}$ of
Einstein-Proca theory, we can perform an inverse Legendre
transform

\ben
\mathfrak{L}(S_{ab},F_{ab})=\bar{\mathfrak{L}}(\mathfrak{g}^{ab},F_{ab})
+\frac{1}{2}\mathfrak{g}^{ab}S_{ab}\,,
\een

to obtain the Lagrangian of the corresponding symmetric affine theory.

\pagebreak

We require that in the weak field limit ($A_a\to 0$, $F_{ab}\to
0$), the Ampere tensor and the Faraday tensor are approximately
equal, i.e. $G_{ab}\approx F_{ab}$. This requirement guarantees
that the theory gives Maxwell equations with a Proca term as the
weak field limit. This imposes restrictions on possible
Lagrangians and it fixes overall normalization of the Lagrangian
density.

In the weak field limit, equation \reeq{ep2} can be rewritten as
\linebreak $A^b{}_{;a;b}-\Box  A_a=-m_V^2A_a$, which gives

\ben
 -\Box A_a+m_V^2A_a-R_{ca}A^c=0\,.
\een

This is the expected equation governing a vector field of mass
$m_V^2$.

In general it is quite difficult to perform the Legendre transform
explicitly. As an example, we consider the Lagrangian suggested by
Eddington

\ben
\mathfrak{L}=-\sqrt{|\det R_{ab}|}\,.
\een

(The sign and normalization are determined by the requirement on
weak field limit mentioned above.)

For this Lagrangian we obtain metric density

\ben
\mathfrak{g}^{ab}=-\sqrt{|\det
R_{ab}|}R^{(ab)}\,,
\een

and Ampere tensor

\ben
\mathfrak{G}^{ab}= \sqrt{|\det R_{ab}|}R^{[ab]}\,,
\een

where $R^{ab}$ is defined as the transposed inverse of $R_{ab}$.

Although it does not seem to be possible to obtain an explicit
formula for $\bar{\mathfrak{L}}$ in terms of standard functions,
we can eliminate $R_{ab}$ to get the relation between the Faraday
and the Ampere tensor (constitutive equation)

\ben
F_{ad}(\delta^d_b+G^d{}_eG^e{}_b)=G_{ab}\,[|\det(\delta^p_q+G^p{}_q)|]^{\frac{1}{n-2}}\,.
\een

In the case of $\mathfrak{L}=-\sqrt{|\det R_{ab}|}$, we get for
the cosmological constant

\ben
\Lambda=-\frac{n-2}{2}\,.
\een

\pagebreak

\section{Symmetric Affine Theory and Scalar Field}

In the framework of symmetric affine theories, we can easily
accommodate bosonic matter fields. We assume that the Lagrangian
density depends on the connection and also on matter fields
and their derivatives. By analogous derivation to that above, it can be
shown that symmetric affine theory with matter fields is
equivalent to
 Einstein-Proca theory with additional matter fields.
In general however, the matter fields are not minimally coupled.

An interesting Lagrangian depending on the connection and scalar
filed was considered by Kijowski \cite{Kijowski1,Kijowski2}

\ben \mathfrak{L}=\frac{1}{V(\varphi)}\sqrt{|\det
S_{ab}-\varphi_{,a}\varphi_{,b}|} \,.\qquad \laeq{lagr} \een

We assume for simplicity that there is only one scalar field, but
all results carry over straightforwardly to the case of $N$ scalar
fields. The Lagrangian is independent of $F_{ab}$, which results
in vector field $A_a$ being identically zero.

Performing the Legendre transform of $\mathfrak{L}$ as before, we
arrive at the Lagrangian

\ben
\bar{\mathfrak{L}}(\mathfrak{g}^{ab},\varphi,\varphi_{,a})
=-\sqrt{g}\left(\frac{1}{2}g^{ab}\varphi_{,a}\varphi_{,b}+V(\varphi)\right)\,.
\een

And the field equations can be rewritten as

\ben
R^{\mathrm M}{}_{ab}=-2\frac{\partial\bar{\mathfrak{L}}}
{\partial \mathfrak{g}^{ab} }\,,\qquad \laeq{ek1}
\een

\ben
g^{ab}\varphi_{;a;b}-V'(\varphi)=0\,.\qquad \laeq{ek2}
\een

Equations \reeq{ek1} and \reeq{ek2} are precisely the nonlinear
Einstein-Klein-Gordon equations.
In this case the scalar field turns out to be minimally coupled.
A similar approach leads to Lagrangians for other matter fields
(minimally coupled in the absence of $F_{ab}$). The Lagrangians
that depend on $\Gamma_{a\phantom{a}c}^{\phantom{a}b}$ only
through $R_{(ab)}$ generally result in
$\Gamma_{a\phantom{a}c}^{\phantom{a}b}$ being metric connection.
If on the other hand the Lagrangian depends on
$\Gamma_{a\phantom{a}c}^{\phantom{a}b}$ in some other way than
through $R_{(ab)}$, the presence of matter field gives rise to the
nonmetricity of $\Gamma_{a\phantom{a}c}^{\phantom{a}b}$.

Interestingly, the Lagrangian \reeq{lagr} naturally arises as
a brane action with scalar fields representing transverse
displacement of the brane.

\pagebreak

\section{Spherically Symmetric Solutions}

We use the equivalence of symmetric affine theories and the nonlinear
Einstein-Proca system to investigate static spherically symmetric
solutions in symmetric affine theories.

We assume that the metric and fields are independent of
time and invariant under time reversal.
 As shown above,  the equation
of motion can be derived from the Lagrangian

\ben
\frac{1}{2}R-\bar{L}-\frac{m_V^2}{2}A^aA_a\,.
\een

$\bar{L}$ in general depends on two invariants $F^{ab}F_{ab}$ and
${}^*F^{ab}F_{ab}$ of the \linebreak electromagnetic field. Since $A_a$ is a
vector, the first invariant is a scalar, \linebreak whereas the second is a
pseudoscalar. The time reversal invariance then requires
${}^*F^{ab}F_{ab}=0$.

This allows us to restrict to the Lagrangian of the form

\ben
\frac{1}{2}R-\frac{1}{4}f(F^2)-\frac{m_V^2}{2}A^aA_a\,,
\een

where $-\frac{1}{4}f(F^2)$ represents a nonlinear term which is
assumed to depend only on $F^2=F_{ab}F^{ab}$.
Note that $\frac{1}{4}f(0)$ can be interpreted as the cosmological
constant.
Since symmetric
affine theory gives $m_V^2=(n-1)(n-2)$, we consider only the case
$m_V^2>0$.

Spherical symmetry enables us to put the metric in the form

\ben
ds^2=-e^{2\alpha(r)}dt^2+e^{2\beta(r)}dr^2+r^2(d\theta^2+\sin^2\theta
d\varphi^2)\,.\footnote{We restrict to metric of signature $(-+++)$. Result for general signature can be deduced by analytic continuation. } \een

The vector potential can be put in the form

\ben
A_a=(u(r),v(r),0,0)\,.
\een

The nonlinear Einstein-Proca equations now read

\ben
R_{ab}-\frac{1}{2}Rg_{ab}=T_{ab}\,,
\een

\ben
G^{ab}{}_{;b}+m_V^2A^a=0\,,
\een
where

\ben
G_{ab}=f'(F^2)F_{ab}\,,
\een

\ben
F_{ab}=A_{b,a}-A_{a,b}\,,
\een

\ben
T_{ab}=f'(F^2)F_{ad}F_b{}^d-\frac{1}{4}f(F^2)g_{ab}+m_V^2(A_aA_b-\frac{1}{2}A^2g_{ab})\,.
\een

\pagebreak

The $01$ component  of the Einstein equation reads $m_V^2uv=0$.
$u=0$ leads to $F^2=0$ and so $v=0$. Therefore it suffices to
consider only the case $v=0$.

Then the Einstein equations and the field equations are equivalent to

\ben
\left[r^2e^{-(\alpha+\beta)}w\right]'=m_V^2r^2ue^{\beta-\alpha} \qquad \textrm{where  } w=f'(F^2)u'\,,\qquad \laeq{EE1}
\een

\ben
\alpha'+\beta'=\frac{1}{2}m_V^2u^2re^{2(\beta-\alpha)}\,,\qquad \laeq{EE2}
\een

\ben
[(1-e^{-2\beta})r]'=r^2\left(\frac{1}{4}f(F^2)-\frac{1}{2}F^2f'(F^2)
+\frac{1}{2}m_V^2u^2e^{-2\alpha}\right)\,,\qquad \laeq{EE3}
\een

where $F^2=-2u'^2e^{-2(\alpha+\beta)}$ ('' $'$ " denotes derivative with respect to $r$).

Equation \reeq{EE1} gives

\ben
\left[r^2e^{-(\alpha+\beta)}wu\right]'=r^2e^{\alpha+\beta}\left(-\frac{1}{2}F^2f'(F^2)
+m_V^2u^2e^{-2\alpha}\right)\,.\qquad \laeq{I1}
\een

We will now assume that the matter fields satisfy the weak energy
condition. This imposes constraints on possible functions
$f(F^2).$
The weak energy condition requires that $T_{ab}U^aU^b\geq0$ for
all timelike and null vectors $U^a$. This requires $T^0{}_0\leq
0$, $T^0{}_0-T^i{}_i\leq 0$ for $i=1,2,3$.
In the case we are considering, this translates to

\ben
\frac{1}{4}f(F^2)-\frac{1}{2}F^2f'(F^2)
+\frac{1}{2}m_V^2u^2e^{-2\alpha}\geq0\,,
\een

\ben
-\frac{1}{2}F^2f'(F^2) + m_V^2u^2e^{-2\alpha}\geq0\,.
\een

Thus the weak energy condition implies that
$r^2e^{-(\alpha+\beta)}wu$ is increasing \linebreak with $r$, and
$M(r)=(1-e^{-2\beta})r$ is increasing with $r$.

Suppose spacetime has an event horizon. By spherical symmetry, the
horizon is a surface $r=const$ given by $\beta\to\infty.$ The
event horizon does not present any natural barrier and a freely
falling observer would not be able to tell (based on local
observations) that  he has passed the horizon. Thus the physically
observable quantities are bounded at the horizon.

$g=r^2e^{\alpha+\beta}\sin\theta$ is bounded at the horizon, which
implies that $\alpha \to - \infty$ at the horizon.
$-R=T^a{}_a=F^2f'(F^2)-f(F^2) -m_V^2u^2e^{-2\alpha}$ and $F^2$ are
bounded at the horizon.
Thus we conclude that $u=0$ at the horizon.

\pagebreak

Suppose spacetime has two horizons. As $r^2e^{-(\alpha+\beta)}wu$
is increasing by \linebreak the weak energy condition and $u=0$ at the
horizons, this means that \linebreak $r^2e^{-(\alpha+\beta)}wu=0$ between two
horizons.
Finally $r^2e^{-(\alpha+\beta)}wu=0$ implies $u=0$. ($w=0$ leads
to $u=const$, but $u=0$ at horizons.)

Suppose that spacetime has one horizon and we impose the following
decay at infinity:
We  assume that $u\to 0$, $u'\to 0$, $r^2uu'\to 0$ as
$r\to\infty$. Then we have that  $r^2e^{-(\alpha+\beta)}wu\to 0$
as $r\to\infty$, ($\alpha+\beta$ is increasing and $f(F^2)$ is
bounded). As $r^2e^{-(\alpha+\beta)}wu$ is an increasing function
that vanishes on the horizon and at infinity,
$r^2e^{-(\alpha+\beta)}wu=0$ and hence $u=0$ between the horizon
and infinity.

If there is no horizon and no singularity at $r=0$
we have $r^2e^{-(\alpha+\beta)}wu=0$ at $r=0$,
and as before we may deduce that $r^2e^{-(\alpha+\beta)}wu=0$
and hence $u=0$ between $r=0$ and $\infty$.

We could imagine a solution with several horizons at $r_1<r_2
\ldots <r_k$. If $r_i$, $r_{i+1}$ are two horizons such that $t$
is timelike and $r$ is spacelike between the horizons, then the
preceding analysis shows that $u=0$ between
 $r_i$ and $r_{i+1}$. The analysis above  can be extended even for
 the regions in which $r$ is timelike and $t$ is spacelike and we may
 conclude that $u=0$  in
$[r_1,\, \infty)$.

\section{Conclusions}{

Symmetric affine formulation of General Relativity suggests that
the connection is the fundamental concept whereas metric is only a
derived notion. Presence of matter gives  rise to nonintegrability
and generally even nonmetricity of the connection.

Symmetric affine theory with no additional matter fields is
equivalent to Einstein gravity with nonlinear electrodynamics
including the Proca term. The mass of the vector boson in the Proca term
is $m_V^2=(n-1)(n-2)$ in reduced Planck units.
This equivalence can be exploited to show that if the weak energy
condition is imposed, there are no nontrivial (meaning including
nonzero $F_{ab}$) static spherically symmetric solutions with
horizons and there are no nontrivial static spherically symmetric
solutions without a singularity.
With additional constraints on $f(F^2)$ ( for example $f'(F^2)>0$)
it could be shown that there are no static solutions with
horizons even in general nonspherical case. This is a particular
instance of  the ``no-hair'' theorem in the framework of symmetric
affine theories.

Symmetric affine theory can naturally include bosonic fields.
Symmetric affine theory with scalar fields gives a natural action
for branes.

\pagebreak

}
\section*{Acknowledgements}
This work has been done under supervision of Prof. G. W. Gibbons.
I would like to thank him for his valuable explanations and very
useful insights. I would also like to acknowledge Trinity College
Heilbronn Fund from which I was funded under Trinity College
Summer Research Scheme.

\end{document}